\documentclass[11pt,a4paper]{article}   %,twocolumn
\usepackage{amsmath}
\usepackage{amssymb}
\usepackage{mathrsfs}

\usepackage{rotating}
\usepackage{enumerate}

\usepackage{tikz}

\usepackage[round]{natbib}

\usepackage{soul}

\usepackage[colorlinks=false,urlcolor=blue]{hyperref}

\renewcommand{\exp}{{\rm e}}

\newcommand{\thick}{h}

\newcommand{\eg}{e.g.\ }
\newcommand{\ie}{i.e.\ }

\usepackage{pifont}
\usepackage[normalem]{ulem}
\usepackage{setspace}
\usepackage{xcolor}
\usepackage{marginnote}

\usepackage{xparse}

\NewDocumentCommand{\marcomm}{mooo}
 {% #1 = text 
  \IfValueTF{#4}
  {\marginnote[\hspace{#3}\framebox{\parbox{#4}{\setstretch{1.0}#1}}]{\hspace{#3}\framebox{\parbox{#4}{\setstretch{1.0}#1}}}[#2] }
  {\IfValueTF{#3}
  {\marginnote[\hspace{#3}\framebox{\parbox{40pt}{\setstretch{1.0}#1}}]{\hspace{#3}\framebox{\parbox{40pt}{\setstretch{1.0}#1}}}[#2] }
  {\IfValueTF{#2}
  {\marginnote[\framebox{\parbox{40pt}{\setstretch{1.0}#1}}]{\framebox{\parbox{40pt}{\setstretch{1.0}#1}}}[#2] }
  {\marginnote[\framebox{\parbox{40pt}{\setstretch{1.0}#1}}]{\framebox{\parbox{40pt}{\setstretch{1.0}#1}}}[0pt] }
  }}}
 
\usepackage[colorinlistoftodos,textwidth=48pt,textsize=scriptsize]{todonotes}

\begin{document}
\title{Wave attenuation and dispersion due to floating ice covers}
\author{L.~J.~Yiew$^{1}$\footnote{
Corresponding author's email: \texttt{lucas\_yiew@tcoms.sg}.
%and \texttt{lucas.yiew@gmail.com}.
Current address: Technology Centre for Offshore and Marine, Singapore (TCOMS), Singapore.
}, 
S.~M.~Parra$^{2,3}$,
D.~Wang$^{3}$, 
D.~K.~K.~Sree$^{4}$,
A.~V.~Babanin$^{5}$
and
A.~W.-K.~Law$^{1,4}$
\\
%{\footnotesize
%$^{1}$}
%\\
{\footnotesize
$^{1}$Environmental Process Modelling Centre, NEWRI, Nanyang Technological University, Singapore}
\\
{\footnotesize
$^{2}$American Society of Engineering Education, DC, United States}
\\ 
{\footnotesize
$^{3}$U.S. Naval Research Laboratory, Stennis Space Center, MS, United States}
\\ 
{\footnotesize 
$^{4}$School of Civil and Environmental Engineering, Nanyang Technological University, Singapore}
\\
{\footnotesize 
$^{5}$Department of Infrastructure Engineering, University of Melbourne, VIC, Australia}
} 

\date{\today}
\maketitle

\begin{abstract}
Experiments investigating the attenuation and dispersion of surface waves in a variety of ice covers are performed using a refrigerated wave flume. 
The ice conditions tested in the experiments cover naturally occurring combinations of continuous, fragmented, pancake and grease ice.
Attenuation rates are shown to be a function of ice thickness, wave frequency, and the general rigidity of the ice cover. 
Dispersion changes were minor except for large wavelength increases when continuous covers were tested.
Results are verified and compared with existing literature to show the extended range of investigation in terms of incident wave frequency and ice conditions.
\end{abstract}

%%%%%%%%%%%%%%%%%%%%%%%%%%%%%%%%%%%%%%%
%%%%%%%%%%%%%%%%% INTRO %%%%%%%%%%%%%%%%%%
%%%%%%%%%%%%%%%%%%%%%%%%%%%%%%%%%%%%%%%

\section{Introduction} \label{sec:intro}

The accelerated melting and weakening of our sea ice cover has contributed to an increase in wave activity in the Arctic \citep{ThoRog14} and Antarctic \citep{Masetal18,Stoetal18}, and is a driver for further losses of ice in both regions.
In response to the retreat of the ice cover, the Arctic has seen an increase in human and maritime activities \citep{Steetal11,Meletal16}.
These recent developments have been motivations for a growing body of research on wave-ice dynamics in order to improve near and long-term projections of ice conditions. 

Particular attention has been directed towards understanding how ocean waves attenuate and disperse as they propagate through partially ice-covered seas,
as is evident by a large collection of works \citep[\eg][]{Wadetal88,WanShe10b,Mosetal15}.
The attenuation and dispersion of waves has been attributed to mechanisms such as wave scattering, viscous damping or hysteresis losses within the ice cover (due to flexure and fracture mechanics), viscous damping under the ice, and inelastic collisions between ice floes \citep{Squetal95}.
The focus of this paper is on quantifying the rate of attenuation and dispersion changes as waves propagate through different types of ice covers.

There have been several efforts to investigate wave damping in ice covers through experimental and theoretical means.
\citet{MarKau81} produced grease ice in a wave tank and measured the decay in wave amplitude.
They reported that attenuation rates were proportional to the square of the wave steepness $ka$, where $k$ is the wavenumber and $a$ is the wave amplitude.
\citet{NewMar97} also investigated the attenuation and dispersion relation of surface waves in two samples of grease ice,
and compared their experimental data with a mass-loading model and a one-layer viscous model. 
The viscous model gave better agreement with the data,
while the mass-loading model was inappropriate for grease ice as it could not capture the increase in wavelengths.
\citet{NewMar99} compared the same data with a two-layer model (an upper viscous and a lower inviscid layer) by \citet{Kel98}.
The Keller model, with a prescribed grease ice viscosity of 2--3$\times$10$^4$ of seawater gave best agreement with their experimental data.
The Keller model was also validated by \citet{WanShe10a}, who performed   experiments similar to \citeauthor{NewMar97}, but on a mixture of grease and pancake ice.
The model was unable to capture the rollover in wavenumbers and the larger attenuation rates in low frequencies.
\citet{ZhaShe15} extended the range of investigation by conducting similar tests on three types of ice covers: frazil/pancake ice, pancake ice and fragmented ice.
Their data was used to validate the viscoelastic model of \citet{WanShe10b}.
In the \citeauthor{WanShe10b} model, the ice cover is homogenised and modelled as viscoelastic layer floating atop an inviscid body of water.
\citeauthor{ZhaShe15} obtained estimates of the effective viscosity $\nu$ and shear modulus $G$ of each ice cover by inversely solving the \citeauthor{WanShe10b} model for $\nu$ and $G$.

Laboratory experiments on wave damping have also been performed using floating synthetic covers. 
\citet{Sutetal17} measured wave attenuation in latex and polypropylene covers,
and \citet{Sreetal18} measured the attenuation and dispersion in covers produced using PVC and PDMS polymers.
In the latter, covers of varying viscoelastic properties and thicknesses were prepared, and their material properties were measured in-situ using a rheometer.
\citeauthor{Sreetal18} also performed a validation of the \citet{WanShe10b} model, and found poor model-data agreement in attenuation rates for the thicker covers in short wavelengths.
This mirrors \citeauthor{Sutetal17}, who also found larger errors between measured attenuation rates for thicker covers in high frequencies and theoretical rates which consider additional damping due to air-water drag and boundary layer effects at the side walls and bottom of the flume.
\citeauthor{Sreetal18} attributed the discrepancy to the viscous dissipation that occurs in an oscillating laminar boundary layer.

In this paper, we present results from a unique set of laboratory experiments where  wave attenuation and dispersion changes are measured for a larger subset of ice conditions than those previously reported.
In our experiments, a refrigerated wave flume is used to generate up to 6 types of ice covers of varying physical and mechanical properties.
Section~\ref{sec:experiments} describes the methods, instrumentation, ice conditions and range of wave conditions tested.
In Section~\ref{sec:attenuationdispersion}, attenuation rates and dispersion relations of waves in each ice cover are presented and compared with existing data.
Section~\ref{sec:discussion} provides a summary of our findings and a discussion on areas of further research.

%%%%%%%%%%%%%%%%%%%%%%%%%%%%%%%%%%%%%%%
%%%%%%%%%%%%%%% EXPERIMENTS  %%%%%%%%%%%%%%%
%%%%%%%%%%%%%%%%%%%%%%%%%%%%%%%%%%%%%%%

\section{Experiments} \label{sec:experiments}

%%% SETUP
\subsection{Setup}

Laboratory experiments investigating wave attenuation and dispersion were implemented using the Sea-Ice-Wind-Wave-Interaction (SIWWI) facility at the University of Melbourne, Australia.
The facility consists of a 14\,m long, 0.75\,m wide wave flume housed within a refrigerated room.
Temperatures within the room can be regulated between ambient temperature to approximately $-12^{\circ}$C.
A variety of ice covers can be produced by varying temperature and wave conditions.

\begin{figure}[h!]
\begin{center}
\includegraphics[width=1\textwidth]{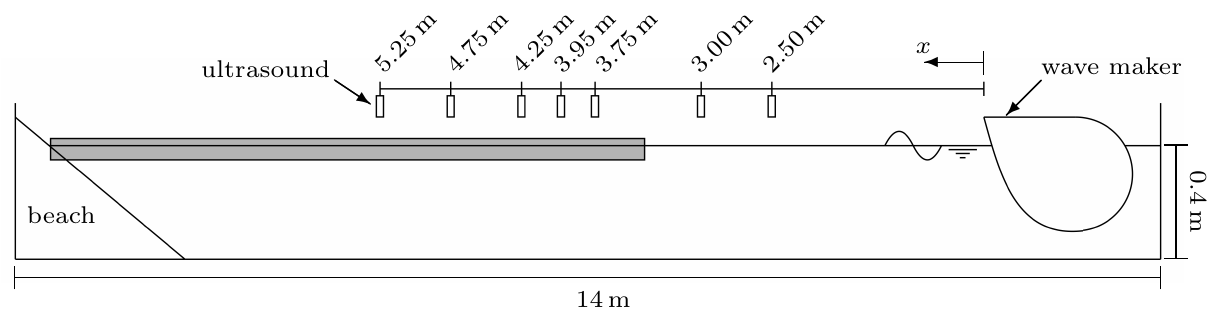}
\caption{
Schematic profile view of the SIWWI wave flume, showing ultrasound sensor placements, and the location of the ice cover (shaded region).}
\label{fig:flume}
\end{center}
\end{figure}

Figure~\ref{fig:flume} shows a schematic profile of the wave flume,
which was filled with freshwater to a depth of 0.4\,m.
During the experiments, monochromatic waves were generated by a piston-type wave maker with an elliptical profile, at the front end of the flume.
A beach at the far end of the flume dissipates incident waves.
In the absence of an ice cover (\ie open-water condition), waves reflected off the beach were measured to contain less than 10\% of the incident wave energy.

%%% FIG %%%

The coordinate $x$ is used to denote the horizontal locations along the length of the flume, with the origin at the wave maker, and the positive direction pointing towards the beach.
Throughout the experiments, an ice-free (open-water) region was maintain around the wave maker and between $x = 0$ to $x \approx 3.375$\,m

An array of 7 ultrasound sensors (General Acoustics ULS40D, USS325) were used to measure the open-water and ice-covered surface profiles along the flume, at a sampling frequency of 50\,Hz.
Each ultrasound sensor has an accuracy of $\pm1$\,mm.
Sensors were positioned 300\,mm above the mean water surface and along the longitudinal centerline of the flume.
Figure~\ref{fig:flume} shows the placement of the sensors and their corresponding $x$-coordinates.
For brevity, ultrasound sensors are referred to as US$_x$, with the subscript $x$ denoting each sensor's $x$-coordinate.
For example, the frontmost sensor is denoted as US$_{2.50}$ and the rearmost sensor is US$_{5.25}$.

The locations of each ultrasound sensor was chosen in order to increase the resolution of results near the ice edge, where attenuation is most clearly observed.
Sensor locations were also selected to avoid interference from waves reflected off the beach.
The first two ultrasound sensors, US$_{2.50}$ and US$_{3.00}$, measured the open-water surface profiles, while the remaining five sensors, from US$_{3.75}$ to US$_{5.25}$, measured the ice-covered surface profiles over a distance of 1.5\,m.
The leading edge of the ice cover was approximately halfway between US$_{3.00}$ and US$_{3.75}$.

Temperature probes were used throughout the experiments to record temperatures within the refrigerated room.
Two temperature probes (one submerged in the flume, and one exposed to air) were used to monitor air and water temperatures during ice production and wave tests.
The probes were placed towards the front of the flume, near the wave maker.

Profile-view videos were also recorded for qualitative purposes.
A videocamera, centerred about the leading ice edge, was used to confirm wave amplitudes and record the presence of overwash, \ie waves which wash over the surface of the ice cover.
Tests with overwash were not analysed as overwashed waves interfered with ice surface profile measurements.

%%% ICE CONDITIONS
\subsection{Ice conditions} \label{subsec:icecdn}

Ice covers of various physical and mechanical properties were produced by varying the wave conditions and air temperatures within the refrigerated room.
Table~\ref{tab:icetypes} summarises the ice conditions considered in these experiments. 

%%% TAB %%%
\begin{table}[htbp]
  \centering
    \begin{tabular}{cc}
    \hline \hline
    Type of ice cover & Variable properties\\
    \hline
    Continuous cover & 3 thicknesses: $\thick = 0.5$, 1 and 1.5\,cm\\
    Fragmented cover & Only 1 condition\\
    Grease ice & 2 concentrations: $c = 30$\% and 40\% \\
    Grease-pancake ice & Only 1 condition\\
    Wide pancake ice & Only 1 condition\\
    Cemented pancake ice & Only 1 condition\\    
    \hline \hline    
    \end{tabular}%
  \caption{Summary of ice conditions tested.}    
  \label{tab:icetypes}%
\end{table}% 

\begin{figure}[h!]
\begin{center}
\includegraphics[width=0.9\textwidth]{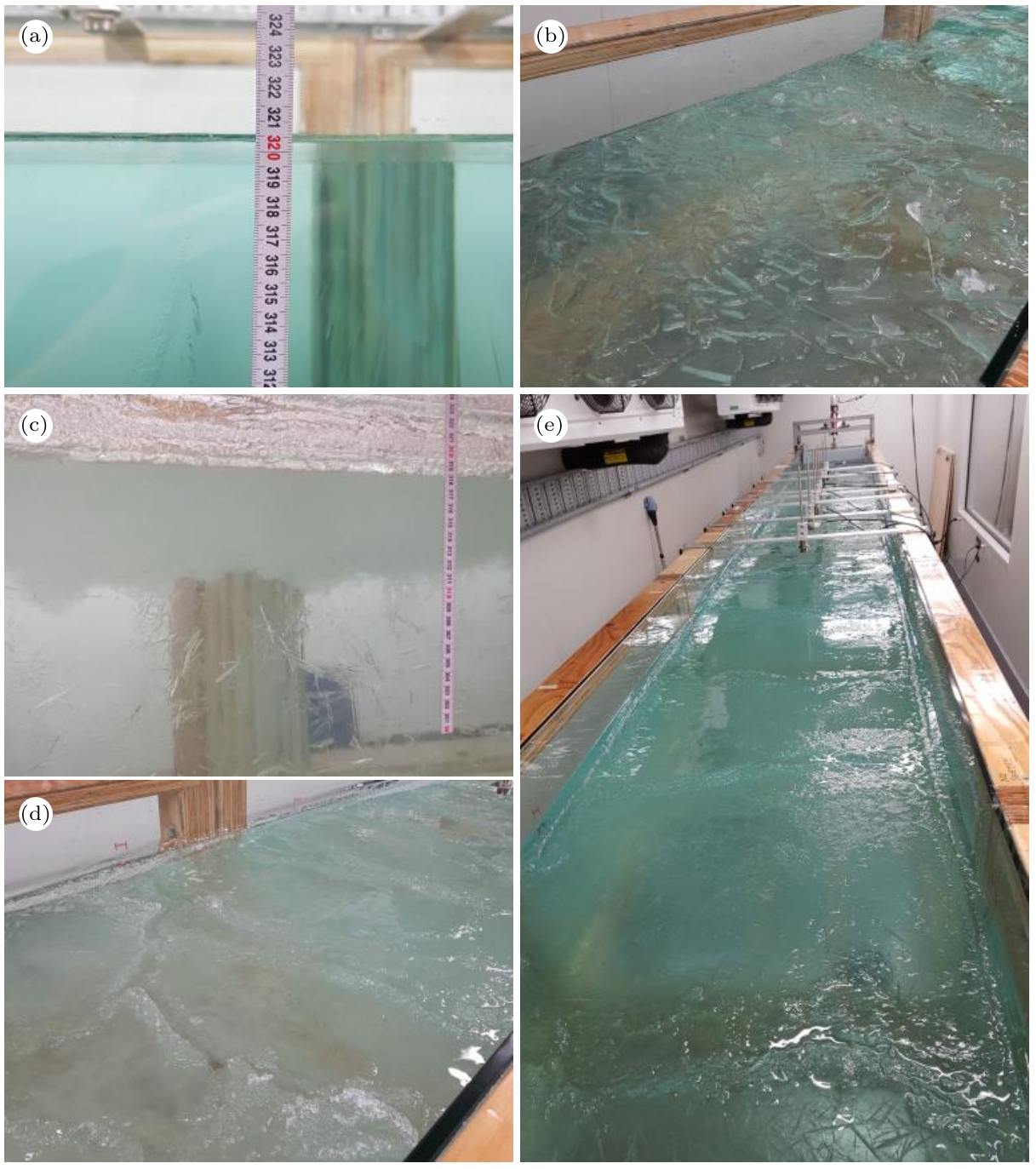}
\caption{
Photos of ice covers tested: (a)~continuous cover, (b)~fragmented cover, (c)~grease ice, (d)~grease-pancakes, and (e)~wide pancakes.}
\label{fig:icetypes}
\end{center}
\end{figure}

In the absence of waves, continuous ice covers of fairly uniform thicknesses were produced by chilling the room overnight (roughly 8--12 hours) at temperatures between $-5$ to $-15^\circ$C.
In these experiments, three continuous covers of thicknesses $\thick = 0.5$, 1 and 1.5\,cm were prepared by varying the room temperatures and freezing durations.
Figure~\ref{fig:icetypes}(a) shows an example of a continuous cover of thickness $t = 1$\,cm.
Before tests were conducted, each cover was detached from the sides of the flume by melting the ice near the side walls using heated strips of metal.
This created a cantilevered sheet of ice and ensured the covers were able to flex freely.
Cantilevers of lengths $L=2.3$, 3.2 and 5\,m were prepared for the 0.5, 1 and 1.5\,cm thick ice sheets, respectively.

%%% FIG %%%

When ice sheets are flexed by waves to the point of failure, angular fragments of ice are formed \citep[see][for example]{Heretal18}.
In the presence of plane waves, ice sheets generally fracture along lines parallel to the wave fronts.
While conducting tests on the thinner $\thick = 0.5$\,cm and 1\,cm continuous covers, wave-induced fracture occurred and resulted in the formation of 3--6 large sheets of ice which were between 0.35\,m to 0.79\,m in length.
Despite the formation of discontinuities in the ice cover, surface profiles were continually measured over a full set of wave conditions.
Due to the length of time (at least one day) required to melt and refreeze the tank, it was not feasible to produce a new ice sheet whenever a fracture occurred.
For the purpose of categorisation, these ice sheets are still referred to as continuous covers.

An ice cover consisting of smaller fragments of ice was also considered in these experiments.
This ice cover was produced by manually breaking a continuous ice sheet into small angular ice fragments of moderately uniform sizes.
Figure~\ref{fig:icetypes}(b) shows an image of the fragmented cover.
The thickness of each individual ice fragment was approximately $1$\,cm, and the overall thickness of the cover (comprising of overlapping layers of ice fragments) was approximately 4\,cm.
The surface areas of each ice fragment ranged from 4.3--675\,cm$^2$, with the mean surface area at 180\,cm$^2$.

Another type of ice cover was created 
using moderately energetic waves and very low temperatures.
\citet{MarKau81} noted that suspensions of ice crystals form and accumulate in cold, windy conditions, in leads and polynyas.
This type of ice is known as grease ice, 
and was produced in SIWWI flume by running waves of $f = 2$\,Hz, $a = 25$\,mm at temperatures between $-8$ to $-12^\circ$C for approximately 3--5 hours.
Figure~\ref{fig:icetypes}(c) shows a profile view of the wave flume with a translucent upper layer of grease ice present.
The properties of grease ice were quantified using its ice concentration,
which is defined as the percentage of mass of ice compared to the overall mass of each suspension.
Ice concentrations were measured by taking the average of 10 samples (approximately 100\,ml each) of grease ice.
Two concentrations, $c = 30\%$ and 40\%, of grease ice were produced for the experiments.

Three types of pancake ice covers were also produced.
The first was produced by compressing and consolidating grease ice using longer, gentler waves of $f = 1.1$\,Hz and $a = 20$\,mm.
Figure~\ref{fig:icetypes}(d) shows an image of the compressed grease-pancakes forming.
Pancakes were approximately 25--30\,cm long and twice as wide.
The overall thickness of the cover was $\thick = 12$\,cm.
The second type of pancake ice was formed by freezing the surface of an ice-free flume at $-10^\circ$C while running waves of $f = 0.7$\,Hz and $a = 10$\,mm.
After approximately 5 hours, rectangular pancakes measuring 40--60\,cm long and spanning the width of the flume were formed.
The overall thickness of the cover was $\thick = 5$\,cm.
Figure~\ref{fig:icetypes}(e) shows an image of these wide rectangular pancakes.
The third type of pancake cover was produced in a similar manner to the wide pancakes, but with slightly shorter waves of $f = 0.8$\,Hz and $a = 10$\,mm.
After 5 hours, round pancakes of 30--40\,cm diameters were produced.
The pancake layer was then frozen overnight at $-2^\circ$C to produce an 8\,cm thick consolidated ice sheet, which was slightly more compliant than the continuous ice cover. 
This ice cover is referred to as a cemented pancake cover.

%%% WAVE CONDITIONS
\subsection{Wave conditions} \label{subsec:wavecdn}

Tests were conducted over a range of regular incident wave frequencies and amplitudes. 
Target incident wave frequencies ranged from $f = 0.8$--1.5\,Hz, with the corresponding incident wavelengths ranging from $\lambda = 0.69$--2.12\,m. Target wave amplitudes ranged from $a = 7.5$--15\,mm.
The resulting wave steepnesses ranged from $ka =  0.022$--0.137.
Table~\ref{tab:testmat} lists the target incident wave conditions used in the experiments.

%%% TAB  %%%
\setlength{\tabcolsep}{0.2cm}
\begin{table} [htbp]
\small
{\begin{minipage}[t][90pt][r]{\textwidth}
\centering
\begin{tabular}{@{\hspace{0pt}} c | c c c c c}
\hline \hline \\[-10pt]
$f$\,[Hz]   & 1.5 & 1.227 & 1.042 & 0.905 & 0.8 
\\[3pt]
$\lambda$\,[m] & 0.69 & 1.03 & 1.39 & 1.76 & 2.12
\\[3pt]
\hline \\[-10pt]
$a$\,[mm] & ~ & ~ & ~ & ~ & ~ 
\\[1pt]
$7.5$ & $\circledast$ & $\ast$ & $\circledast$ & $\ast$ & $\circledast$
\\[3pt]
$10$  & $\circledast$ & $\ast$ & $\circledast$ & $\ast$ & $\circledast$
\\[3pt]
$12.5$  & $\ast$ & $\ast$ & $\ast$ & $\ast$ & $\circledast$
\\[3pt]
$15$  & $\circledast$ & $\ast$ & $\circledast$ & $\ast$ & --
\\[3pt]
\hline \\[-10pt]
$n$ & 20 & 17 & 14 & 10 & 7 \\[3pt]
\hline \hline
\end{tabular} 
\end{minipage}}
\caption{Target incident wave frequencies $f$, wavelengths $\lambda$, amplitudes $a$, and number of waves produced in each wave packet $n$. Asterisks and circles indicate test conditions for the first and second set of tests, respectively. } 
\label{tab:testmat}
\end{table}

For each test, regular incident wave packets were generated.
The duration of each wave packet, \ie number of waves produced, was varied depending on the incident wave frequency (and corresponding wave celerity). 
This was done to minimise the interference from waves reflected off the beach. 
The last row of Table~\ref{tab:testmat} lists the number of waves, $n$, produced for each wave frequency. 

Table~\ref{tab:testmat} shows the target wave conditions for two sets of tests.
The first set of test conditions was used for the 1.5\,cm thick continuous cover, fragmented cover and grease-pancake cover.
The second set of test conditions was used for the remaining ice conditions.
To ensure repeatability of results, each amplitude and frequency combination in the first set of tests was repeated 3 times.
For the second set of test conditions, only the largest amplitudes for each wave frequency were repeated twice.

%%% DATA PROCESSING
\subsection{Data processing} \label{sec:attenuationdispersion}

\begin{figure}[h!]
\begin{center}
\includegraphics[width=1\textwidth]{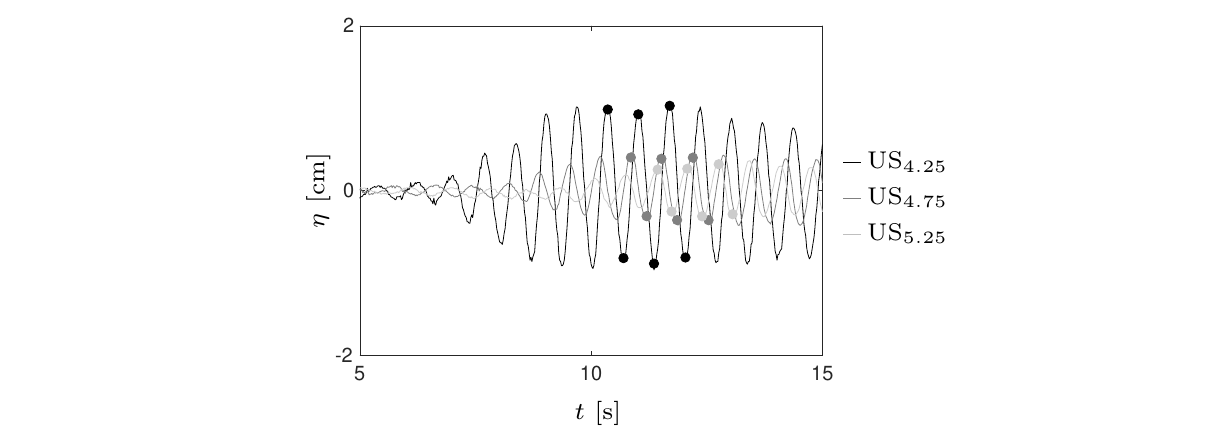}
\caption{
Example of surface profiles $\eta$ over time $t$ at various ultrasound locations. 
Bullets indicate peaks and troughs used in the calculation of wave amplitudes and celerities.
}
\label{fig:surfaceprofiles}
\end{center}
\end{figure}

Figure~\ref{fig:surfaceprofiles} shows an example of the data obtained from the ultrasound sensors.
Surface profiles $\eta$, are plotted as a function of time.
In this example, surface profiles measured at US$_{4.25}$, US$_{4.75}$ and US$_{5.25}$ are shown for a fragmented cover test when incident waves of $f = 1.5$\,Hz and $a = 10$\,mm were generated.

%%% FIG %%%

Wave amplitudes at each ultrasound location are calculated over three wave periods within a steady-state interval.
Bullets in Figure~\ref{fig:surfaceprofiles} indicate the selected peaks and troughs for the example test.
Peaks and troughs at each ultrasound location correspond to the same consecutive wave fronts.
Wave amplitudes are calculated as half the average peak to trough distance over the selected three-wave-period interval.
For tests with repeated wave conditions, wave amplitudes are also averaged with respect to repeated tests.

Wave celerities along the flume are derived using the same peaks and troughs used to calculate wave amplitudes.
Celerities between each adjacent ultrasound sensor are calculated as the quotient of the distance between each sensor
and the mean time difference of matching peaks and troughs between each sensor.
Again, for repeated tests, means are calculated over all repeated tests.

Open water (incident wave) celerities are calculated using the time differences between US$_{2.50}$ and US$_{3.00}$,
while ice-covered celerities are calculated using the time differences between adjacent ultrasounds from US$_{3.75}$ to US$_{5.25}$.
Mean ice-covered celerities are taken as the average of the four celerities calculated between US$_{3.75}$ and US$_{5.25}$.

\begin{figure}[h!]
\begin{center}
\includegraphics[width=0.9\textwidth]{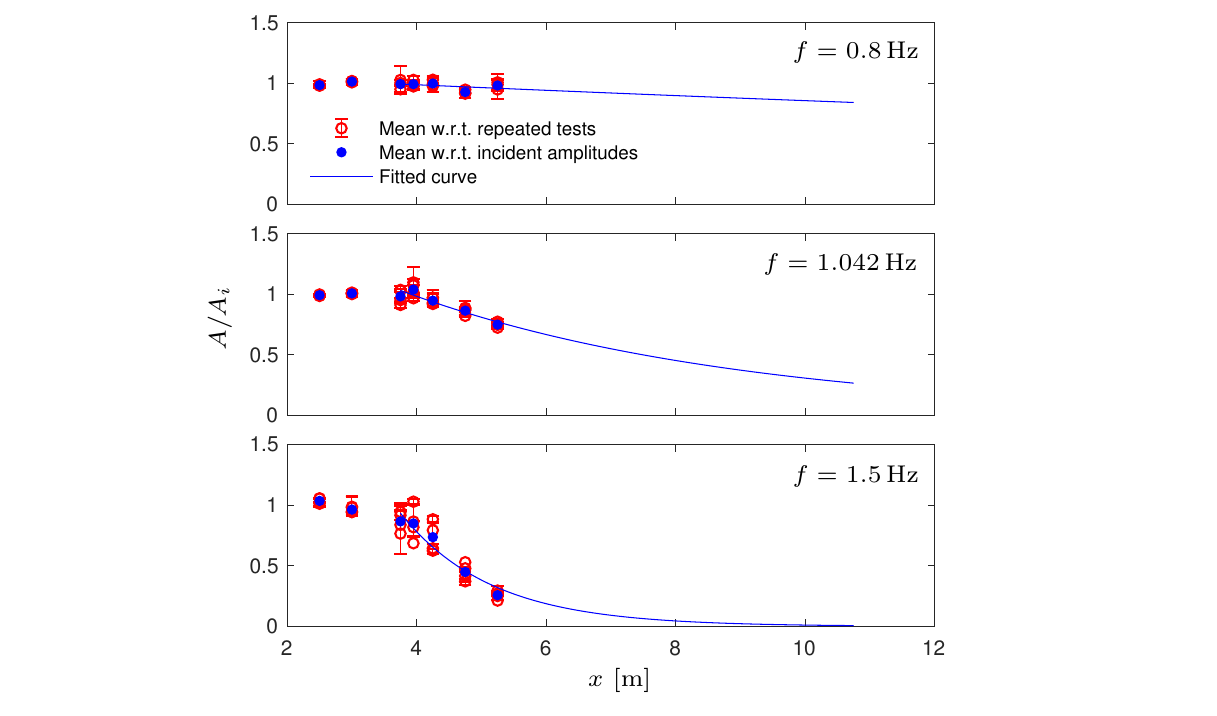}
\caption{
Measured wave amplitudes $A$ normalised with respect to the measured incident amplitude $A_i$, for the fragmented ice cover tests, as a function of horizontal distance $x$.
Each panel represents data for a specific wave frequency $f$.
}
\label{fig:waveamp}
\end{center}
\end{figure}

Figure~\ref{fig:waveamp} shows the measured wave amplitudes $A$, at each ultrasound location for the fragmented cover tests.
Data in the top, middle and bottom panels are grouped according to incident wave frequencies $f = 0.8$, 1.042 and 1.5\,Hz, respectively.
All wave amplitudes in Figure~\ref{fig:waveamp} are normalised with respect to the measured incident wave amplitude $A_i$, which is calculated as the mean of amplitudes at US$_{2.50}$ and US$_{3.00}$.

%%% FIG %%%

In each panel, red circles represent mean amplitudes of repeated tests (\ie tests using the same incident wave frequencies and amplitudes),
and the corresponding error bars indicate the maximum and minimum amplitudes for each wave condition.
Mean amplitudes are also calculated with respect to all incident wave amplitudes tested.
These means (indicated by blue bullets in Figure~\ref{fig:waveamp}) are obtained by averaging the means of repeated tests (red circles).

Mean amplitudes with respect incident amplitudes are then used to obtain best-fit curves and attenuation rates.
Best-fit curves are derived by fitting the means (blue circles) from US$_{3.75}$ to US$_{5.25}$ to the equation
\begin{equation}
\label{eq:attenuation}
A(x) = A_0 \exp^{-\alpha (x-x_0)}
\end{equation}
via the MATLAB function \texttt{fit}.
In Equation~(\ref{eq:attenuation}), $\alpha$ and $A_0$ are the fitted parameters, with $\alpha$ denoting the attenuation rate, and $A_0$ being the amplitude at $x_0$, which is the $x$-coordinate of US$_{3.75}$.
In Figure~\ref{fig:waveamp}, fitted curves are extrapolated to highlight the change in attenuation rates between different wave frequencies.

%%%%%%%%%%%%%%%%%%%%%%%%%%%%%%%%%%%%%%%
%%%%%%%%%%%%%  DATA PROCESSING  %%%%%%%%%%%%%%
%%%%%%%%%%%%%%%%%%%%%%%%%%%%%%%%%%%%%%%
\section{Results}

%%% ATTENUATION
\subsection{Attenuation}

\begin{figure}[h!]
\begin{center}
\includegraphics[width=1\textwidth]{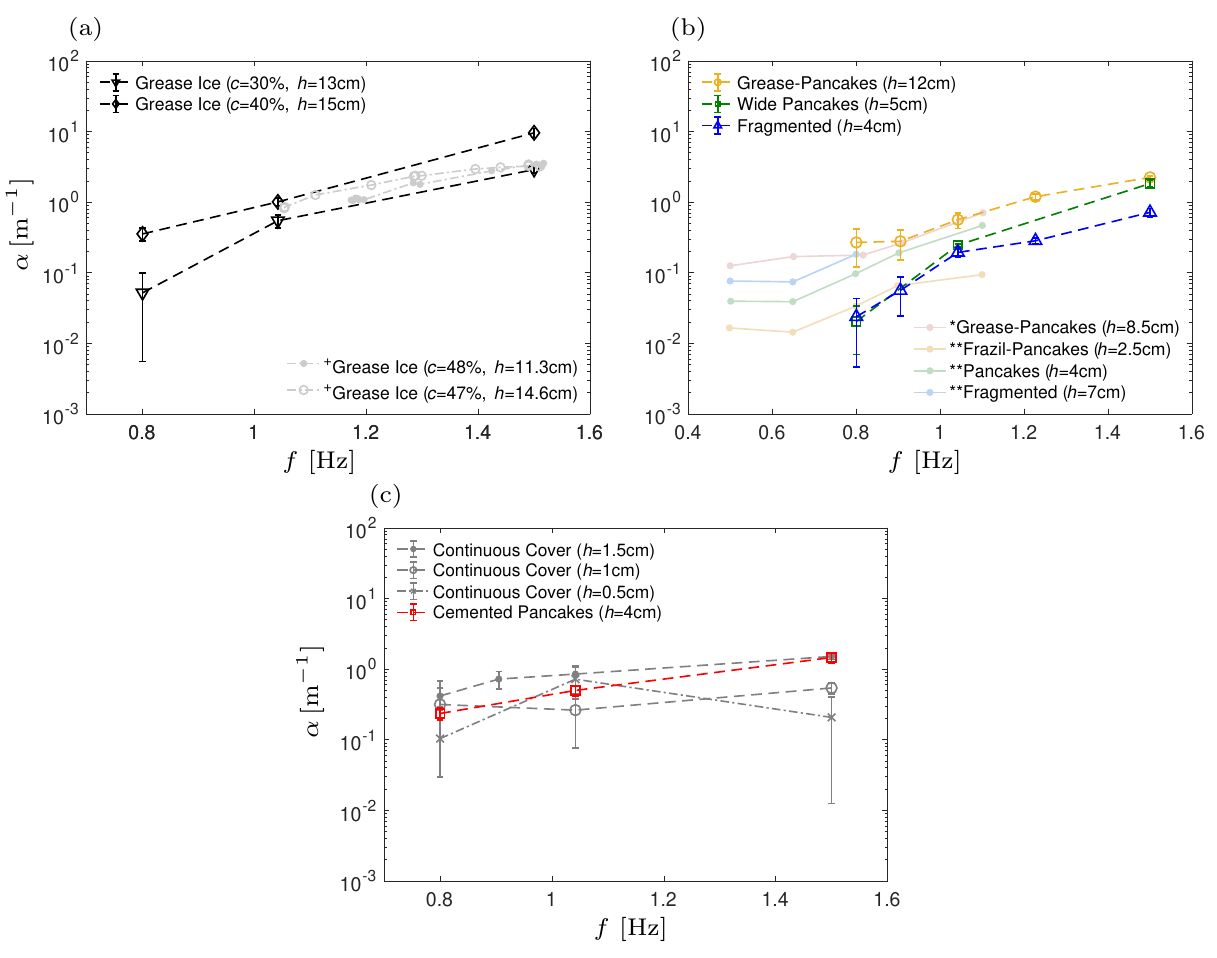}
\caption{
Attenuation rates $\alpha$ for each ice condition, as a function of wave frequency $f$.
Properties such as overall thicknesses of each ice cover $h$, and ice concentrations $c$ (for grease ice), are indicated in the legend.
Data from $^+$\citet{NewMar97}, $^*$\citet{WanShe10a} and $^{**}$\citet{ZhaShe15} are also shown for comparison.
}
\label{fig:attenuation}
\end{center}
\end{figure}

Attenuation rates for each ice condition are plotted as a function of wave frequency in Figure~\ref{fig:attenuation}.
Error bars indicate the 50\% confidence interval for the fitted parameter $\alpha$.
For clarity, results are divided into three groups of ice conditions (top left, top right and bottom panels), according to the general compliance of each ice cover.
Figure~\ref{fig:attenuation}(a) shows the attenuation rates for two concentrations of grease ice, with grease ice being the most compliant cover.
Figure~\ref{fig:attenuation}(b) shows the attenuation rates for grease-pancakes, wide pancakes and the fragmented cover.
The third group of ice covers in Figure~\ref{fig:attenuation}(c) are the least compliant.
This group consist of the cemented pancake cover and three continuous covers of different thicknesses.

%%% FIG %%%

Figures~\ref{fig:attenuation}(a), (b) and (c) show a general trend of attenuation rates increasing monotonically with wave frequency.
The only ice covers which do not have a monotonic increase are the $h=0.5$\,cm and 1\,cm continuous covers.
Attenuation rates also appear to increase with the thickness of each cover.
The only exceptions are for the $h=0.5$\,cm continuous cover and wide pancakes at $f=1.042$\,Hz and 0.8\,Hz, respectively.

Of the nine ice conditions in our experiments, the least overall wave decay occurred when then the fragmented cover was tested.
Attenuation rates for this cover range from $\alpha = 2.37 \times 10^{-2}$ to $7.15 \times 10^{-1}$\,m$^{-1}$.
The fragmented cover's attenuation rates are only slightly larger than others  when $f = 0.8$\,Hz (with $\alpha = 2.03 \times 10^{-2}$\,m$^{-1}$ for wide pancakes),  
and when $f = 1.5$\,Hz (with $\alpha = 2.07 \times 10^{-1}$\,m$^{-1}$ for the $h=0.5$\,cm continuous cover).

The largest overall wave decay occurred when the $c=40$\%, $h = 15$\,cm grease ice cover was tested.
Attenuation rates for this ice cover range from $\alpha = 3.54 \times 10^{-1}$ to $9.57$\,m$^{-1}$.
Although grease ice is relatively compliant, this large wave decay is not completely unexpected as its thickness is up to 2 orders of magnitude larger than other covers.
Attenuation rates for this cover are, on average, 4 times larger than the less concentrated and slightly thinner $c=30$\%, $h=13$\,cm grease ice cover,
with the latter having attenuation rates between $\alpha = 5.18 \times 10^{-2}$ to $2.89$\,m$^{-1}$.
In Figure~\ref{fig:attenuation}(a), attenuation rates from \citet{NewMar97} are also shown for comparison.
\citeauthor{NewMar97} performed similar tests on two grease ice covers, with both covers having approximately equal concentrations ($c\approx48$\%) but slightly different average thicknesses.
Attenuation rates obtained by \citeauthor{NewMar97} are within the range our results.
Both of their grease ice covers also exhibit an increase in attenuation rates with thickness.

Figure~\ref{fig:attenuation}(b) shows the attenuation rates for grease-pancakes, wide pancakes and the fragmented cover.
The wide pancakes and fragmented cover have similar thicknesses ($h = 5$ and 4\,cm, respectively), while the grease-pancake cover is up to 3 times thicker at $h=12$\,cm.
Attenuation rates for wide pancakes range between $\alpha = 2.03 \times 10^{-2}$ to 1.83\,m$^{-1}$, and are, on average, 1.6 times larger than the fragmented cover.
Attenuation rates for these two ice covers deviate with increasing wave frequency, with wide pancakes having attenuation rates an order of magnitude larger than the fragmented cover when $f = 1.5$\,Hz.
Of the three covers in this figure, grease-pancakes have the largest attenuation rates, which range between $\alpha = 2.68 \times 10^{-1}$ to 2.24\,m$^{-1}$.
This is primarily due to it being the thickest cover.

In Figure~\ref{fig:attenuation}(b), attenuation rates from \citet{WanShe10a} and  \citet{ZhaShe15} are overlaid for comparison.
In \citeauthor{WanShe10a}, tests were conducted on a grease-pancake cover, while in \citeauthor{ZhaShe15}, frazil-pancakes, pancakes and a fragmented cover was tested.
(Grease ice is generally defined as a denser accumulation of frazil ice.)
In both studies, tests were conducted over wave frequencies $f=0.5$ to 1.1\,Hz.
Attenuation rates from these experiments show the same increases with respect to cover thickness.
When comparing similar ice conditions and thicknesses between the two studies and our experiments, we see a definitive overlap in results.
For example, attenuation rates for \citeauthor{WanShe10a}'s grease-pancakes are within the range of error of our grease-pancakes between $f=0.8$ to 1.1\,Hz.
The attenuation rates for our grease-pancakes are, however, largest due to our cover being thickest at $h=12$\,cm, compared to $h=8.5$\,cm for \citeauthor{WanShe10a}'s grease-pancakes and $h=2.5$\,cm for \citeauthor{ZhaShe15}'s frazil-pancakes.
Differences in cover thickness also explains the difference in attenuation rates for \citeauthor{ZhaShe15}'s and our fragmented covers.

In Figure~\ref{fig:attenuation}(c), attenuation rates for the cemented pancake cover are compared with the rates for the three continuous covers.
Attenuation rates for the cemented pancake cover range from $\alpha=2.35 \times 10^{-1}$ to 1.46\,m$^{-1}$.
These values are within the overall range of the continuous covers.
Comparisons between the three continuous covers show that the thickest cover, $\thick = 1.5$\,cm, experiences the largest attenuation, with $\alpha$ ranging from $4.15 \times 10^{-1}$ to 1.51\,m$^{-1}$.
Attenuation decreases with decreasing ice thickness,
except for a single outlier for the $\thick = 0.5$\,cm cover at $f = 1.042$\,Hz, where $\alpha = 7.18 \times 10^{-1}$\,m$^{-1}$ is larger than that of the $\thick = 1$\,cm cover, $\alpha = 2.62 \times 10^{-1}$\,m$^{-1}$.

%%% DISPERSION
\subsection{Dispersion}

\begin{figure}[h!]
\begin{center}
\includegraphics[width=1\textwidth]{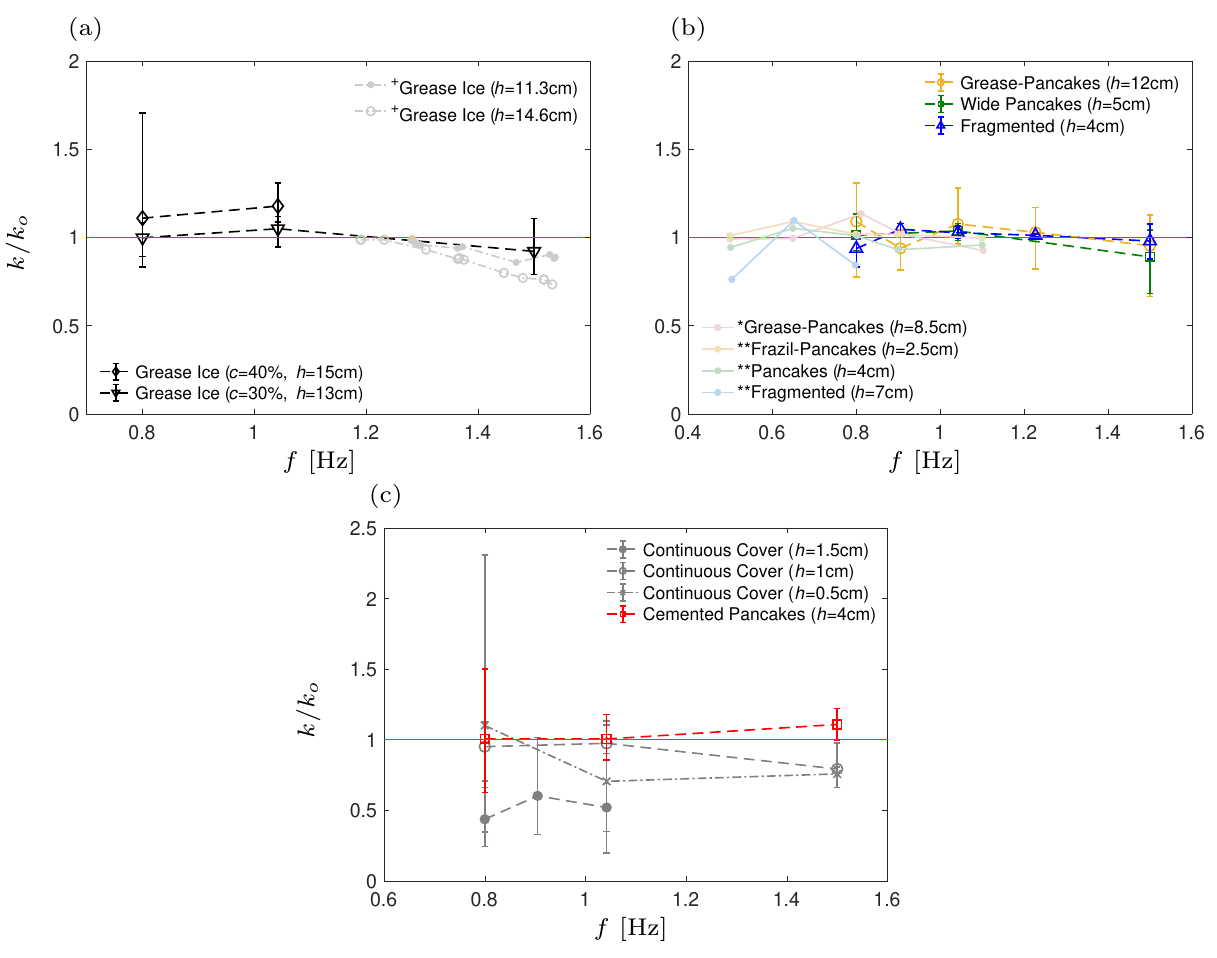}
\caption{
Wavenumbers $k$ of each ice cover in comparison to the open water wavenumber, and expressed as a function of wave frequency $f$.
Data from $^+$\citet{NewMar97}, $^*$\citet{WanShe10a} and $^{**}$\citet{ZhaShe15} are shown for comparison.
Lines, symbols and colours consistent with those in Figure~\ref{fig:attenuation}).
}
\label{fig:wavenumber}
\end{center}
\end{figure}

The relative wavenumber $k$ of each ice cover with respect to the open water wavenumber $k_o$ are shown Figure~\ref{fig:wavenumber} as a function of wave frequency.
Wavenumbers are derived using the measured wave celerities $c_p$ according to the equation $k = 2 \pi f / c_p$.
In Figure~\ref{fig:wavenumber}, errorbars represent the overall maximum and minimum wavenumbers for all tests at each wave frequency.
Results are again separated into three groups of ice covers, as with Figure~\ref{fig:attenuation}.
Data for the $c=40$\% grease ice is limited to $f = 0.8$--1.042\,Hz as surface profiles in higher frequency waves were too small to identify corresponding peaks and troughs.
Data for the $\thick = 1.5$\,cm continuous cover is also limited to the same range as overwash, which was present in higher frequencies, caused difficulties in  identifying surface profiles of the ice cover.
Wavenumbers measured in the absence of an ice cover (\ie open water, and referred to as $k_o$) are also shown for comparison.
Data points greater than $k/k_o = 1$ represent a decrease in wave celerity (and wavelength) as waves propagate through the ice cover.

%%% FIG %%%

Wavenumbers for ice conditions in Figures~\ref{fig:wavenumber}(a) and (b) are generally close to the open water condition.
The fragmented cover has the smallest absolute mean difference of 1.5\%,
while the $c=40$\% grease ice has the largest absolute mean difference of 9.4\%.
Wide pancakes have the largest deviation from $k_o$ at $f = 1.5$\,Hz, with the difference of 1.13\,m$^{-1}$ corresponding to a 12.5\% decrease in wavenumber and a 14.3\% increase in wavelength.

In Figure~\ref{fig:wavenumber}(c), wavenumbers for the $\thick = 0.5$ and 1\,cm continuous covers and cemented pancakes are generally comparable to the open water condition between $f = 0.8$--1.042\,Hz, but deviate from $k_o$ with increasing wave frequency.
Cemented pancakes experience an increase in wavenumber (wave shortening), while all other continuous covers experience wave lengthening between $f = 0.8$ to 1.5\,Hz.
The $\thick = 1.5$\,cm continuous cover has the largest average change in wavenumber, with the largest difference of 2.1\,m$^{-1}$ (when $f = 0.8$\,Hz) corresponding to a 103\% increase in wavelength.

Data from \citet{NewMar97}, \citet{WanShe10a} and \citet{ZhaShe15} are again overlaid for comparison.
In Figure~\ref{fig:wavenumber}(a), we see that wavenumbers for both of \citeauthor{NewMar97}'s grease ice covers are mostly smaller than $k_o$ over their range of frequencies tested ($f = 1.05$ to 1.51\,Hz).
\citeauthor{NewMar97} also found wavelength increases of up to 36.4\%.
In contrast, wavenumbers for our grease ice covers are larger than the open water case, except at $f = 1.5$\,Hz, where the $k/k_o < 1$ for the $c=30$\% cover.
The maximum wavelength increase for our grease ice covers was much smaller than \citeauthor{NewMar97}'s, at 6.4\%.
Wavenumbers for \citeauthor{NewMar97}'s thicker cover are also smaller than their thinner cover.
This differs from our results, which show larger wavenumbers for our thicker (and more concentrated cover) as compared to our thinner, less concentrated cover. 

Wavenumbers from \citeauthor{WanShe10a} and \citeauthor{ZhaShe15} have better agreement with our data for the ice covers in Figure~\ref{fig:wavenumber}(b).
Wavenumbers for these ice covers are much closer to $k_o$.
The maximum difference between \citeauthor{WanShe10a} and \citeauthor{ZhaShe15}'s wavenumber and $k_o$ is just under 24\%.
This compares well our results, which show a maximum difference of 8\% from $k_o$.

%% VISCOSITY
\subsection{Grease ice viscosity}

A separate set of tests were conducted to measure the viscosity of a sample of grease ice.
The viscosity of the $c=40$\% grease ice was measured by conducting a simple falling sphere experiment.
Three glass spheres, measuring 17, 25 and 39\,mm in diameter and weighing 7, 20 and 79\,g, respectively, were allowed to fall through a clear plastic tube containing a volume of grease ice.
A volume of polydimethylsiloxane (PDMS), with known physical properties, was also used as a reference fluid.
The time taken for each sphere to fall 0.5\,m was recorded.
Tests were repeated 10 times for each sphere, and the mean velocities of each sphere in grease ice and PDMS ($\overline{V}_\text{g}$ and $\overline{V}_\text{p}$, respectively) were used to calculate the viscosity of grease ice, $\mu_{\text{g}}$ according to the equation,
\begin{equation}
\mu_{\text{g}} = 
\frac
{ (\rho_{\text{s}} - \rho_{\text{g}}) / \overline{V}_{\text{g}} }
{ (\rho_{\text{s}} - \rho_{\text{p}}) / \overline{V}_{\text{p}} }
\,
\mu_{\text{p}} ,
\end{equation}
with $\rho_\text{s}$ denoting the density of the glass sphere, $\rho_\text{g} = 900$\,kg\,m$^{-3}$ the density of grease ice, $\rho_\text{p} = 965$\,kg\,m$^{-3}$ the density of PDMS, and $\mu_\text{p} = 4.6$\,Pa.s the viscosity of PDMS.

A dynamic viscosity $\mu_\text{g} = 0.12 \pm 0.05$\,Pa.s was obtained in this experiment.
This value is two orders of magnitude smaller than those reported by \citet{NewMar99}, who obtained grease ice viscosities ranging from 22--32\,Pa.s by tuning viscous fluid models to fit laboratory measured wave decay and wavenumber changes due to a grease ice cover.

\begin{figure}[h!]
\begin{center}
\includegraphics[width=1\textwidth]{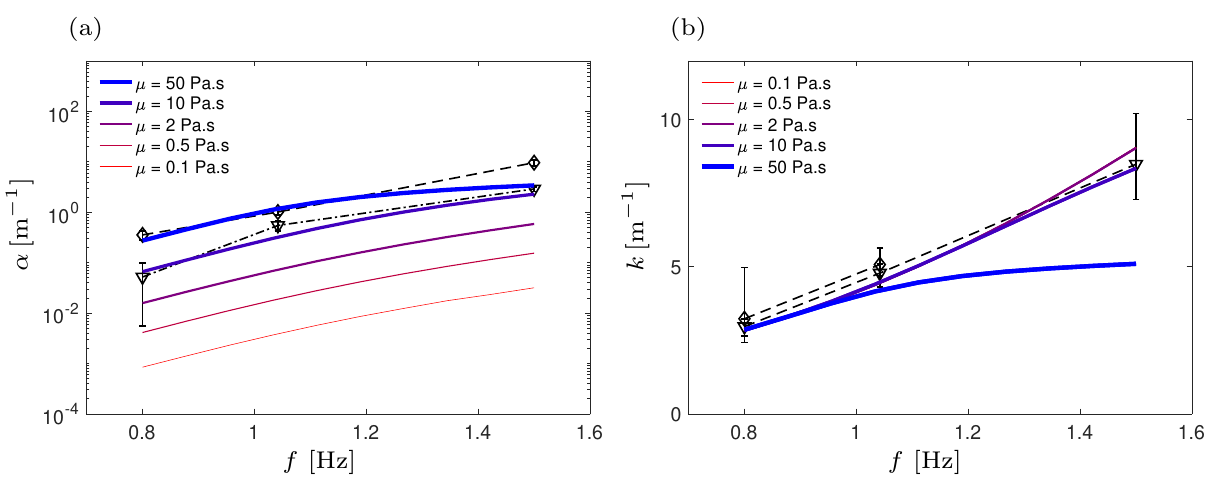}
\caption{
(a) Attenuation rates $\alpha$ and (b) wavenumbers $k$ predicted by the \citet{WanShe10b} model for a range of viscosities $\mu$, in comparison to experimental results for the  $c = 30$\% and 40\% grease ice covers (lines and symbols for grease ice data are consistent with those in Figures~\ref{fig:attenuation} and \ref{fig:wavenumber}).
}
\label{fig:sensitivity}
\end{center}
\end{figure}

Using these range of values for $\mu$, a sensitivity study was conducted to compare experimental data with theoretical predictions using the model of \citet{WanShe10b}.
In this model, an ice cover is introduced by defining values for viscosity and rigidity.
For grease ice, it is common to assume rigidity is zero.
Figure~\ref{fig:sensitivity} shows the attenuation rates and wavenumbers given by the model in comparison to data for the $c = 30$\% and 40\% grease ice covers.
Theoretical results for five values of $\mu$, between 0.1 to 50\,Pa.s, are shown.
The attenuation data clearly shows that the range of viscosities obtained from the falling sphere experiment are too small.
Viscosities between $\mu = 10$--50\,Pa.s give attenuation rates which better fit the experimental data/
However wavenumbers obtained when $\mu = 50$\,Pa.s deviate from the experimental data, especially in higher frequencies, as shown in Figure~\ref{fig:sensitivity}(b).
Viscosities of $\mu \approx 10$\,Pa.s are needed to give reasonable agreement for both attenuation rates and wavenumbers.

%%% FIG %%%

%%%%%%%%%%%%%%%%%%%%%%%%%%%%%%%%%%%%%%%
%%%%%%%%%%%%%% CONCLUSIONS %%%%%%%%%%%%%%%%
%%%%%%%%%%%%%%%%%%%%%%%%%%%%%%%%%%%%%%%

\section{Conclusions}\label{sec:discussion}

Laboratory experiments using a refrigerated wave flume were performed to quantify wave attenuation and dispersion in a variety of freshwater ice covers. 
The process of generating these ice covers, as well as the physical properties of the prepared covers are documented in this paper.

Attenuation and dispersion (in terms of wavenumber changes) were measured for 6 types of ice covers, including continuous and fragmented covers, and combinations of pancake and grease ice covers.  
9 ice conditions were also tested, including different thicknesses and concentrations, in the case of grease ice.

Wavenumbers for the continuous cover had the largest difference compared to the open water wavenumber.
Wavenumbers for the continuous cover were generally less than the open water case, indicating an increase in wavelength.
The thickest continuous cover had the largest increase in wavelength of 103\%.

Attenuation was shown to increase monotonically with wave frequency in all except one case. 
The apparent rollover in attenuation rates at high wave frequencies (observed reported by \citealt{Kohetal11} and \citealt{Lietal17}) did not occur in our experiments.
Comparisons between similar types of ice showed that attenuation rates also increase with ice cover thickness.

Attenuation rates also vary according to the overall rigidity of the ice cover, with the more rigid but thin continuous and cemented pancake covers having larger attenuation rates than the less rigid and similarly thin fragmented and pancake covers.
However, grease ice, which is generally considered to be the least rigid ice cover, had the largest overall attenuation rates, when compared with other covers of similar thicknesses.

Viscous dissipation within the ice cover is a definite source of wave attenuation, particularly for grease ice. 
However viscous dissipation within the ice-water boundary layer can also contribute to further attenuation. 
Experiments by \citet{Sreetal18} suggests this to be the case, especially in higher wave frequencies.
Additional tests measuring boundary layer turbulence were performed concurrently with the experiments reported here, however the results of those tests are outside the scope of this paper and will be duly reported in a subsequent publication.

%%% Acknowledgements %%%

\section*{Acknowledgements}

L.J.Y., D.K.K.S. and A.W.-K.L. acknowledge funding support from the U.S. Office of Naval Research Global.
A.V.B. acknowledges support of the DISI Australia-China Centre through grant ACSRF48199 and the U.S. Office of Naval Research through grant N00014-17-1-3021.
The authors thank Filipo Nelli, Alberto Alberello, Jason Monty, and Erick Rodgers for technical support during the experiments, 
and Hayley Shen for her comments on the manuscript.
\\[10pt]

\bibliographystyle{apa}      % References
%\bibliography{Bibli}

\end{document}